\documentclass[aps,showpacs,prb,floatfix,twocolumn,citeautoscript]{revtex4-1}
\usepackage{graphicx}
\usepackage{bm,amsmath,amssymb,mathrsfs,dcolumn}
\usepackage{subfigure}
\usepackage{units}
\usepackage{hyperref}
\usepackage[usenames]{color}
\hypersetup{pdfborder=0 0 0,colorlinks=true,citecolor=blue,linkcolor=blue}
\input epsf

\newcommand{\BN}{\textit{h}-BN}
\newcommand{\MBN}{\textit{h}-BN}
\newcommand{\MS}{MoS$_2$}
\newcommand{\SL}{MoS$_2$}

\begin{document}

\title{Controlling the Schottky barrier at \MS{}$|$metal contacts by inserting a BN monolayer} 

\author{Mojtaba Farmanbar}
\author{Geert Brocks}
\email{g.h.l.a.brocks@utwente.nl}
\affiliation{Faculty of Science and Technology and MESA$^{+}$ Institute for Nanotechnology, University of Twente, P.O. Box 217, 7500 AE Enschede, The
Netherlands}

\begin{abstract}
Making a metal contact to the two-dimensional semiconductor \MS{} without creating a Schottky barrier is a challenge. Using density functional calculations we show that, although the Schottky barrier for electrons obeys the Schottky-Mott rule for high work function ($\gtrsim 4.7$ eV) metals, the Fermi level is pinned at 0.1-0.3 eV below the conduction band edge of \MS{} for low work function metals, due to the metal-\MS{} interaction. Inserting a boron nitride (BN) monolayer between the metal and the \MS{} disrupts this interaction, and restores the \MS{} electronic structure. Moreover, a BN layer decreases the metal work function of Co and Ni by $\sim 2$ eV, and enables a line-up of the Fermi level with the \MS{} conduction band.  Surface modification by adsorbing a single BN layer is a practical method to attain vanishing Schottky barrier heights.
\end{abstract}

\date{\today}
\maketitle

{\color{red}\it Introduction.} Single layers of transition metal dichalcogenides (TMDs) such as molybdenite, \MS{}, can be exfoliated through micromechanical cleavage, similar to graphene.\cite{Geim:nat13} In contrast to graphene however,  a \MS{} monolayer is a semiconductor with a sizable band gap of 1.8 eV,\cite{Mak:prl10} which has triggered a large interest in TMD semiconductor devices.\cite{Radisavljevic:nnano11,Wang:nnano12,Wang:nanol12}  Contacting \MS{} to metal electrodes remains a problem, as it tends to produce unexpectedly high contact barriers and resistances. Early photoemission experiments claimed that the Schottky barriers at \MS{}$|$metal interfaces obey the ideal Schottky-Mott rule,\cite{Lince:prb87} suggesting the possibility to control the Schottky barrier height (SBH). In particular, the SBH for electrons might be reduced to zero using a metal with a sufficiently low work function. However, the more recent device experiments do not give zero SBHs, neither for metals with high work functions, nor for metals with low work functions.\cite{Liu:acsnano12,Das:nanol13,ChenJR:nanol13,Fontana:scirep13,Kaushik:apl14,Kang:apl14}

SBHs of metal contacts with conventional semiconductors such as Si often only weakly depend on the metal species, and the Fermi level is pinned inside the semiconductor band gap.\cite{Tung:apr14} Common models used to explain Fermi level pinning rely upon having a strong (chemical) interaction at the metal-semiconductor interface that yields a large density of interface states with energies in the semiconductor band gap. Unlike Si, \MS{} has no dangling bonds at its surface. Its interaction with metal surfaces should therefore be relatively weak, which makes it rather unlikely that midgap interface states are formed at a high density. Indeed a recent density functional theory (DFT) study claims there is only a weak Fermi level pinning at \MS{}$|$metal interfaces.\cite{Gong:nanol14} That still leaves the prospect of a zero SBH using a metal that has a sufficiently low work function.

In this paper we study the Schottky barriers at \SL{}$|$metal interfaces by DFT calculations with the objective of designing a contact with zero SBH. We start from a series of metals covering a wide range of work functions (3.8-5.8 eV). Unlike previous studies we do not so much focus on the chemical interactions with specific metals,\cite{Popov:prl12,Chen:nanol13,Kang:prx14} but on establishing general rules for the SBHs. We show that for clean \MS{}$|$metal interfaces with high work function metals the Fermi level is not pinned in the \MS{} band gap, and the SBH shows Schottky-Mott behavior.\cite{Tung:apr14} This breaks down for low work function metals, and the Fermi level gets pinned just below the \SL{} conduction band, leading to a finite SBH in the range 0.1-0.3 eV. The metal-\MS{} interaction at the interface perturbs the electronic structure of \MS{}, its conduction bands in particular, creating a density of interface states just below the conduction band that pins the Fermi level. 

We ``unpin'' the Fermi level by inserting a \BN{} monolayer between the metal surface and \MS{}. It breaks the direct metal-\MS{} interaction and destroys the interface states. Like graphene, \MS{} is physisorbed on the \MBN{}-covered substrate,\cite{Bokdam:nanol11,Gillen:prb14} which leaves its electronic structure nearly unperturbed. Moreover, adsorption of \BN{} on a metal surface commonly decreases the work function considerably. It turns high work function metals such as Co and Ni into low work function substrates. The combined effects of breaking the metal-\MS{} interaction and lowering the metal work function yields zero SBHs for contacts between \MBN{}-covered Co or Ni and \MS{}.

{\color{red}\it Computational details.}
We use projector augmented waves (PAW) as implemented in the VASP code.\cite{Kresse:prb93,Blochl:prb94b,Kresse:prb96,Kresse:prb99} The \SL{}$|$metal interface is modeled by a slab of four layers of metal atoms with a monolayer \MS{} adsorbed on one side. We optimize all atomic positions, keeping the layer of metal atoms furthest removed from the adsorbant in its bulk geometry. We force the metal lattice to be commensurable to the \MS{} lattice, choosing in-plane supercells with a mismatch between the \MS{} and the metal lattices of less than 1\%. The metal-adsorbant binding distance is important for obtaining an accurate interface potential profile, and in some cases this distance depends sensitively on the DFT functional.\cite{Bokdam:prb14b} The PBE generalized gradient approximation\cite{Perdew:prl96} and the optB88-vdW-DF van der Waals density functional (vdW-DF)\cite{Dion:prl04,Klimes:prb11} give \MS{}$|$Ag, Au, Pd and Pt(111) binding distances within 0.1 \AA\ of one another and interface potential steps within 0.05 eV. The local density approximation (LDA)\cite{Perdew:prb81} yields on average $\sim 0.2$ \AA\ shorter binding distances and $\sim 0.15$ eV larger potential steps. We use PBE in the following to avoid the risk of overbinding commonly found with LDA. Whereas the interface potential step does not depend too critically on the DFT functional, the work function of a clean metal surface $W_\mathrm{M}$ can be more sensitive,\cite{Bokdam:prb14a,Gong:nanol14} which is then reflected in the calculated SBHs, see Eq.~(\ref{eq:sb}).

\begin{figure}
\includegraphics[width=8.5cm]{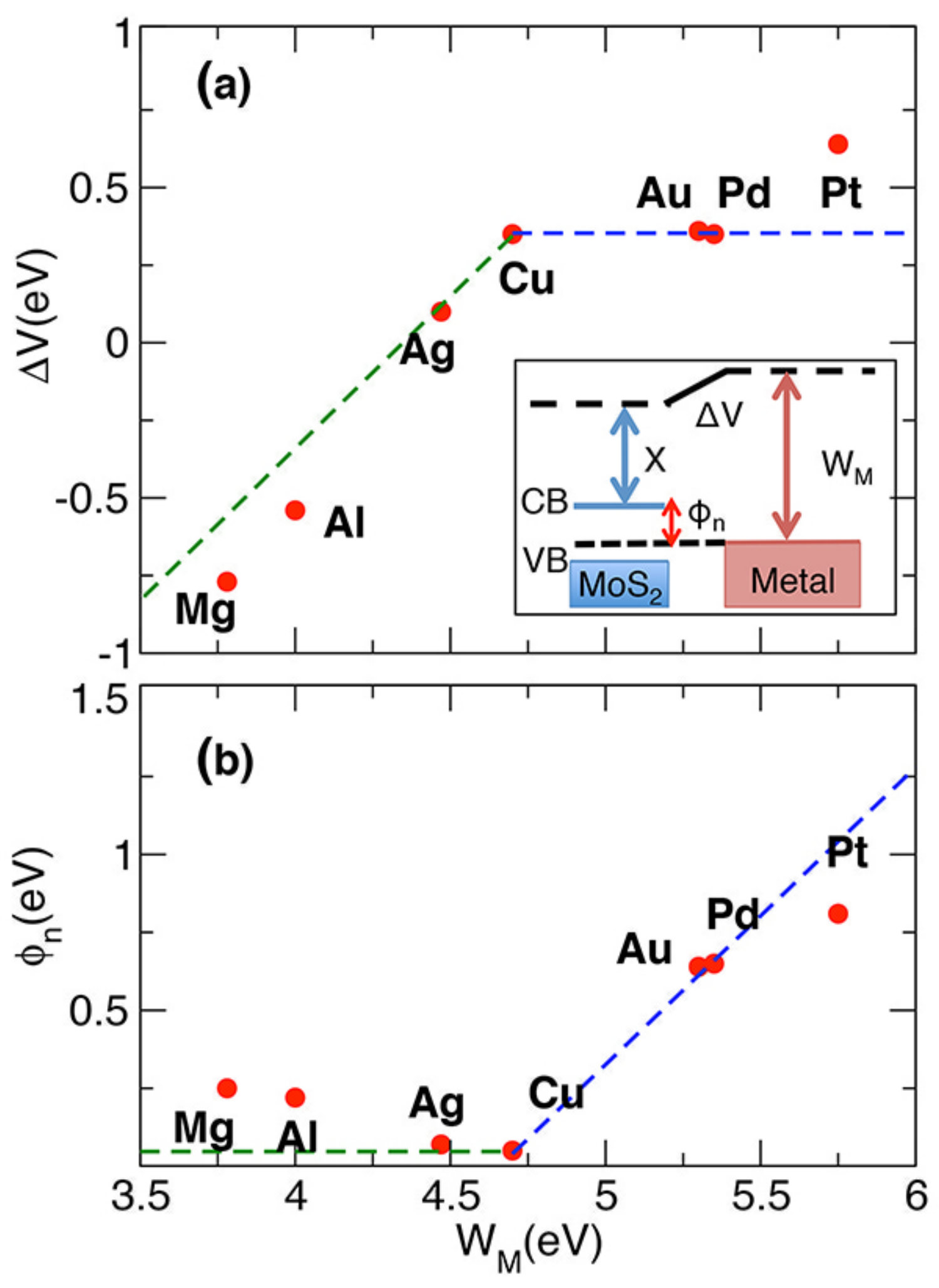}
\caption{(Color online) (a) The potential step $\Delta V$ at the \SL{}$|$metal interface versus $W_\mathrm{M}$, the work function of the clean metal surface. (b) The Schottky barrier height (SBH) for electrons $\Phi_n$. The blue and green dashed lines in (a) and (b) indicate the  Schottky-Mott rule and Fermi level pinning, respectively. Inset: schematic energy diagram of the interface, with $\chi$ the electron affinity of \SL{}.
}
\label{fig:deltav}
\end{figure}

{\color{red}\it \MS{}$|$metal interfaces.}
The SBH for electrons can be written as
\begin{equation}
\Phi_\mathrm{n} = W_\mathrm{M} - \chi - \Delta V, 
\label{eq:sb}
\end{equation}
with $W_\mathrm{M}$ the work function of the clean metal surface, $\chi$ the electron affinity of \SL{}, and $\Delta V$ the potential step formed at the \SL{}$|$metal interface, see Fig. \ref{fig:deltav}. The potential step can be calculated without resorting to the details of the potential profile across the interface, or its electronic structure, because $\Delta V = W_\mathrm{M} - W_\mathrm{ads|M}$, where $W_\mathrm{ads|M}$ is the work function of the metal surface covered with \SL{}. The results are shown in Fig. \ref{fig:deltav}(a). The SBH is then obtained from Eq. (\ref{eq:sb}), using the calculated $\chi = 4.30$ eV, cf. Fig. \ref{fig:deltav}(b). 

Two regimes can be distinguished in Fig. \ref{fig:deltav}(a). Metals with $W_\mathrm{M}\gtrsim 4.7$ eV yield a similar $\Delta V$, whereas $\Delta V$ strongly depends on $W_\mathrm{M}$ for $W_\mathrm{M} \lesssim 4.7$ eV. The cross-over between the two regimes occurs for Cu, where $W_\mathrm{Cu} \approx \chi + \Delta V$ and the SBH is minimal, see Fig. \ref{fig:deltav}(b). The results for the high work function regime are consistent with the common observation that physisorption results in a net decrease of the work function, i.e., a positive $\Delta V$. This has been explained in terms of the Pauli exchange repulsion between the electrons of the metal and those of the overlayer yielding a net pushback of electrons into the metal.\cite{Bokdam:prb14b} The effect is fairly moderate for \SL{}, resulting in $\Delta V\approx 0.35$ eV for the high work function metals, and a size that does not depend critically on the details of the metal or the interface structure. 

With a constant $\Delta V$ the SBH simply follows the work function, i.e., the slope $S = d\Phi_\mathrm{n}/d W_\mathrm{M} \approx 1$, which is the Schottky-Mott rule. This rule is typically found in the absence of any interface states with energies in the semiconductor band gap, which is consistent with the \MS{}$|$metal interaction being weak. If ideal Schottky-Mott behavior would persist for the low work function metals, the SBH would vanish for $W_\mathrm{M} < W_\mathrm{Cu}$. Clearly this is not the case in Fig. \ref{fig:deltav}(b). The SBH has a minimum at Cu, but it increases again for the low work function metals. 

\begin{figure}
\includegraphics[width=8.5cm]{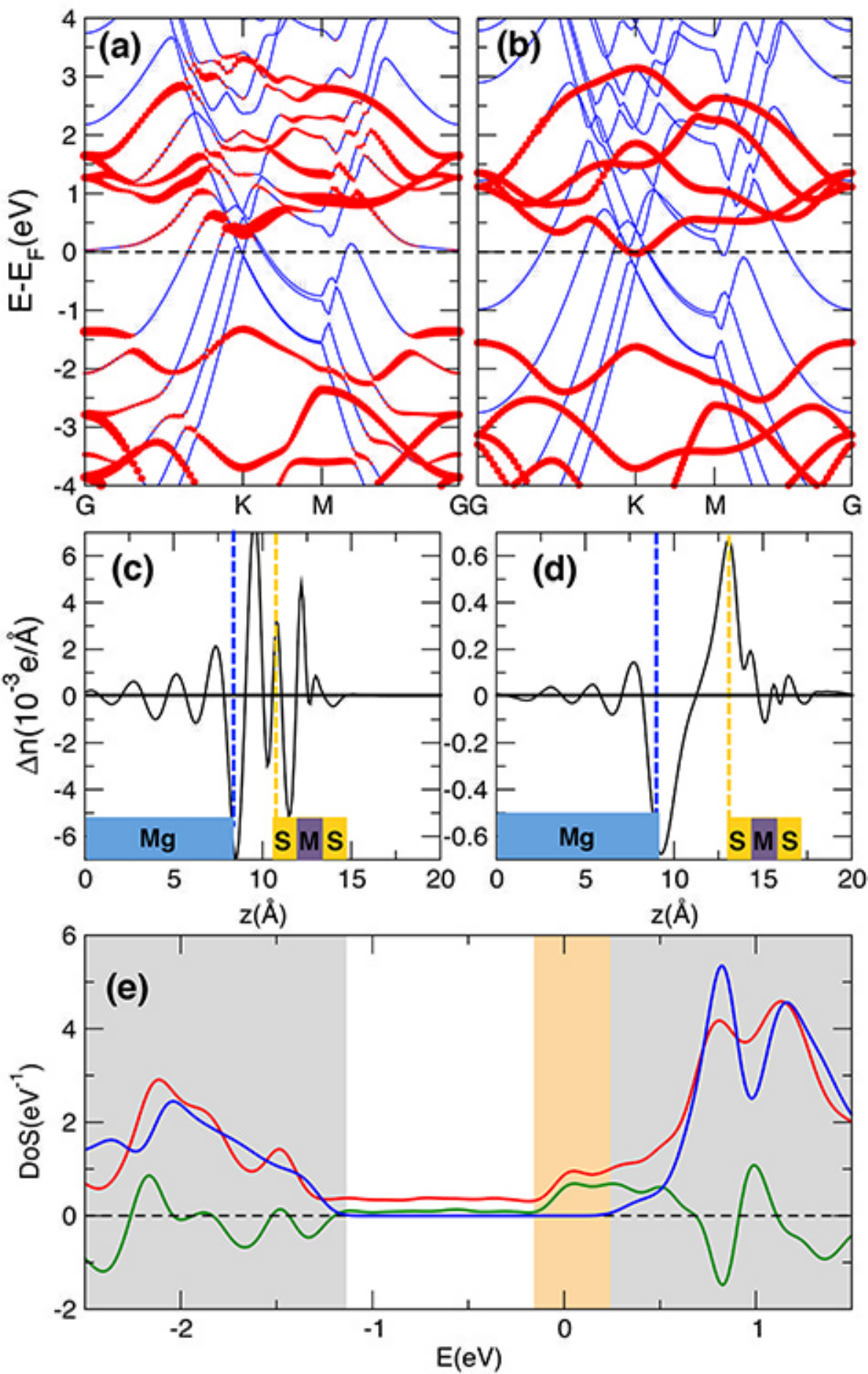}
\caption{(Color online) (a) Band structures of the \SL{}$|$Mg(0001) interfaces at the equilibrium distance $d_\mathrm{eq}=2.2$ \AA\ and (b) at $d=5$ \AA. The red color measures a projection of the wave function on the \MS{} orbitals. The Fermi level is set at zero energy. (c,d) The electron difference density $\Delta n(z)$, corresponding to the interface distances of (a,b), respectively. (e) The total density of states (red) corresponding to (a) $\mathrm{DoS}_\mathrm{ads|Mg}$, (blue) of free-standing \SL{} $\mathrm{DoS}_\mathrm{ads}$, and (green) the difference $\Delta\mathrm{DoS} = \mathrm{DoS}_\mathrm{ads|Mg} - \mathrm{DoS}_\mathrm{ads} - \mathrm{DoS}_\mathrm{Mg}$. The orange shading indicates the position of the gap states, created by the interaction at the interface.
}
\label{fig:mgbands}
\end{figure}

We analyze this behavior for Mg, being the metal with the lowest work function in this study. The band structure of the $(1\times 1)$ \SL{}$|$Mg(0001) interface is shown in Fig. \ref{fig:mgbands}(a). At the optimized equilibrium distance $d=2.2$ \AA, the \MS{} bands are significantly perturbed by the interaction with the substrate. This perturbation can be visualized by comparing the density of states (DoS) of \SL{}$|$Mg to that of free-standing \SL{}, see Fig. \ref{fig:mgbands}(e). The \MS{}$|$Mg interaction leads to interface states in the \SL{} band gap that are energetically close to the bottom of the \SL{} conduction band. This is seen most clearly in the difference DoS, i.e. $\Delta\mathrm{DoS} = \mathrm{DoS}_\mathrm{ads|M} - \mathrm{DoS}_\mathrm{ads} - \mathrm{DoS}_\mathrm{M}$, represented by the green curve in Fig. \ref{fig:mgbands}(e), where $\mathrm{DoS}_\mathrm{ads|M},\mathrm{DoS}_\mathrm{ads},\mathrm{DoS}_\mathrm{M}$ are the DoSs of the interface, the free-standing adsorbate, and the clean metal substrate, respectively. 

The $\Delta\mathrm{DoS}$ is negligible for energies inside the \SL{} band gap, except for a region $\lesssim 0.4$ eV below the conduction band edge (shaded orange in Fig. \ref{fig:mgbands}(e)). The negative sign of the calculated potential step at the \MS{}$|$Mg interface, $\Delta V = -0.77$ eV, shows that electronic charge is transferred from the Mg substrate to the \MS{} overlayer. These electrons populate the interface states, thereby pinning the Fermi level in the band gap. For high work function metals the Fermi level is well within the \SL{} gap, where the density of interface states is negligible. The corresponding SBHs then obey the Schottky-Mott rule. It is only for low work function metals, when the Fermi level approaches the bottom of the \SL{} conduction band, that interface states become noticeable and pin the Fermi level.

One can destroy the interface states by breaking the \MS{}$|$Mg interaction. This is demonstrated by Fig. \ref{fig:mgbands}(b), which gives the band structure of \SL{}$|$Mg(0001) with the adsorbant placed at an artificially large distance $d=5$ \AA\ from the metal surface. The \MS{} bands are unperturbed, as there is no chemical interaction with the substrate at this distance. The Fermi level is in the \MS{} conduction band, which is not surprising as the work function of Mg(0001), $W_\mathrm{Mg}=3.8$ eV is much smaller than the \SL{} electron affinity $\chi=4.3$ eV. The calculated interface potential step, $\Delta V_\mathrm{ni} = -0.5$ eV, indeed corresponds to $W_\mathrm{Mg} - \chi$, as it should for electron transfer from Mg to \SL{} to equilibrate the Fermi level.  It results in a zero SBH, according to Eq. (\ref{eq:sb}).

This interpretation is confirmed by the electron density difference $\Delta n(z) = n_\mathrm{ads|M}(z) - n_\mathrm{ads}(z) - n_\mathrm{M}(z)$,  where $n_\mathrm{ads|M}(z),n_\mathrm{ads}(z),n_\mathrm{M}(z)$ are the plane-averaged electron densities of the interface, the free-standing adsorbate, and the clean metal substrate, respectively. At an \MS{}$|$Mg distance $d=5$ \AA, $\Delta n(z)$ shows an accumulation of electrons at the position of \SL{}, and an electron depletion at the Mg surface, see Fig. \ref{fig:mgbands}(d), consistent with an electron transfer from Mg to \MS{}, which creates an interface dipole and potential step to equilibrate the Fermi level. At the equilibrium \MS{}$|$Mg distance, $\Delta n(z)$ shows a much more complicated pattern, see Fig. \ref{fig:mgbands}(c), which is consistent with an interface interaction that alters the electronic structure.

\begin{figure}
\includegraphics[width=8.5cm]{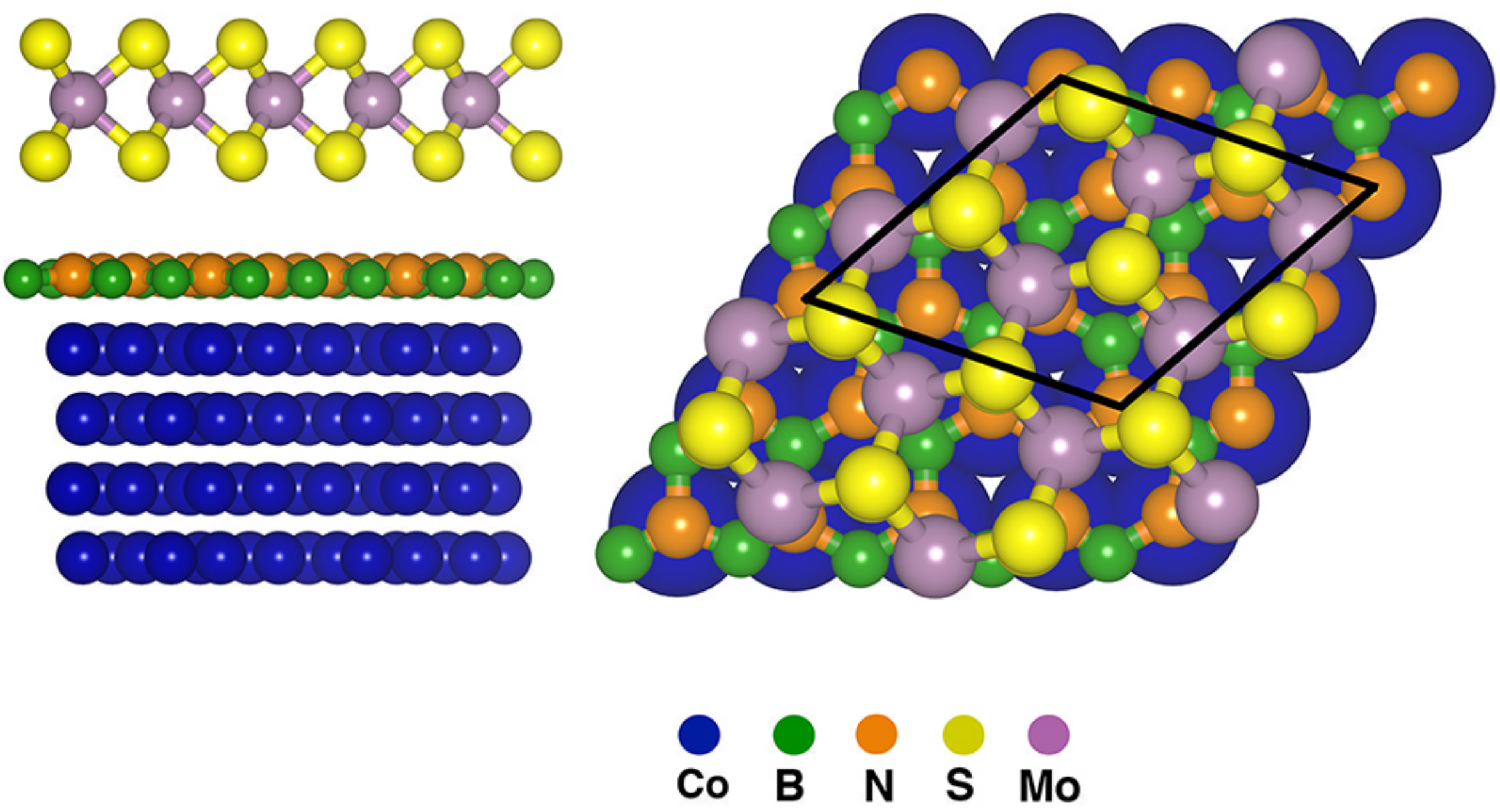}
\caption{(Color online) Top and side view of the \SL{}$|$\MBN{}$|$Co(111) structure. The black rhombus indicates the surface supercell.
}
\label{fig:struc}
\end{figure}

{\color{red}\it \MS{}$|$\MBN{}$|$metal interfaces.}
A Schottky barrier at a \MS{}$|$metal contact is unavoidable, as the interface interaction leads to states that pin the Fermi level below the conduction band. Breaking this interaction by introducing a vacuum spacing at the interface, is not a practical way to make a metal-semiconductor contact. However, inserting an inert layer between the metal and \MS{} can be. This layer has to be sufficiently thin as not to form a high barrier for electron transport. A monolayer of \BN{} is ideally suited. A single layer of \BN{} can be deposited or grown on a range of metal substrates, and it is stable under ambient conditions. Substrates consisting of a transition metal (111) surface covered by a \BN{} monolayer, are widely available. Sandwiching a monolayer of \BN{} between two metal electrodes gives metallic conduction,\cite{Yazyev:prb09} which indicates that the layer is transparent to electrons. 

We build \MS{}$|$\MBN{}$|$metal structures by putting a $(2\times 2)$ \MS{} cell on top of a $(\sqrt{7}\times \sqrt{7})$ \BN$|$Co or Ni cell, see Fig. \ref{fig:struc}, with the \BN{} layer in a $(1 \times 1)$ registry with the underlying metal surface, as in Refs. \onlinecite{Bokdam:prb14a,Bokdam:prb14b}. As before, we fix the lattice constant of \MS{}, and adapt the lattices of \BN{} and the metal(111) substrates accordingly, which requires a 4\% squeeze of the \BN{} lattice. A \BN{} monolayer is chemisorbed on Co and Ni(111) surfaces, but \MS{} and \BN{} are bonded by a weak, van der Waals, interaction. Such interactions are not represented in the PBE functional, so we use the optB88-vdW-DF van der Waals density functional here. 

Adsorption of \MBN{} has a dramatic effect on the work function; it reduces the work functions of Co and Ni(111) by 1.9 eV and 1.8 eV, respectively. These reductions result from large interface dipoles that are formed at the \BN{}$|$metal interfaces, where Pauli exchange repulsion between the electrons at the interface gives an important contribution.\cite{Bokdam:prb14b}  Adsorbants in the form of self-assembled monolayers (SAMs) are commonly proposed in order to modify substrate work functions. However, SAMs often suffer from disorder, which diminishes their effect. Adsorption of \BN{} leads to a well-defined structure that is much less susceptible to disorder, and gives a sizable work function lowering. Moreover \MBN{} presents a surface that not only is chemically relatively inert, but also does not change its structure upon adsorbing further layers. 

\begin{figure}
\includegraphics[width=8.5cm]{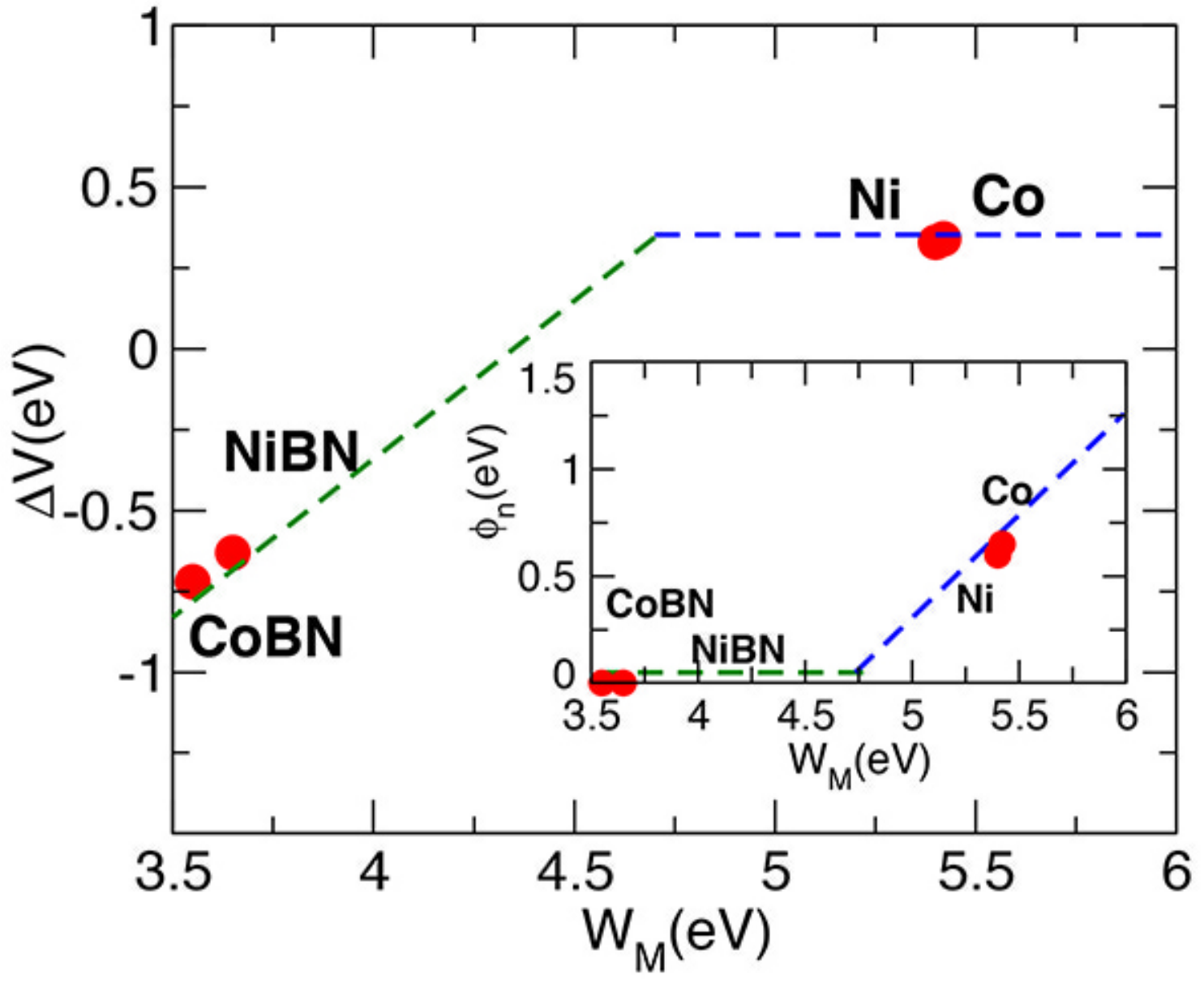}
\caption{(Color online) $\Delta V$ versus $W_\mathrm{M}$. Inset: $\Phi_n$ versus $W_\mathrm{M}$. The blue and green dashed lines indicate the Schottky-Mott rule and Fermi level pinning, as in Fig. \ref{fig:deltav}. 
}
\label{fig:bn}
\end{figure}

The implications of work function lowering by \MBN{} adsorption are clearly demonstrated in Fig. \ref{fig:bn}. Direct adsorption of \SL{} on Co and Ni(111) gives a behavior that is typical for high work function metals. The potential step at the \MS{}$|$metal interface is $\Delta V \approx 0.35$ eV, and the SBHs follow the Schottky-Mott rule. In contrast, adsorption of \SL{} on \MBN{}$|$Co and Ni(111) substrates gives a negative $\Delta V$, and it gives a zero SBH. Inserting a \BN{} layer has not only effectively decreased the substrate work function, but it has also weakened the \MS{}$|$substrate interaction that yielded Fermi level pinning and nonzero SBHs for clean low work function metal substrates.

\begin{figure}
\includegraphics[width=8.5cm]{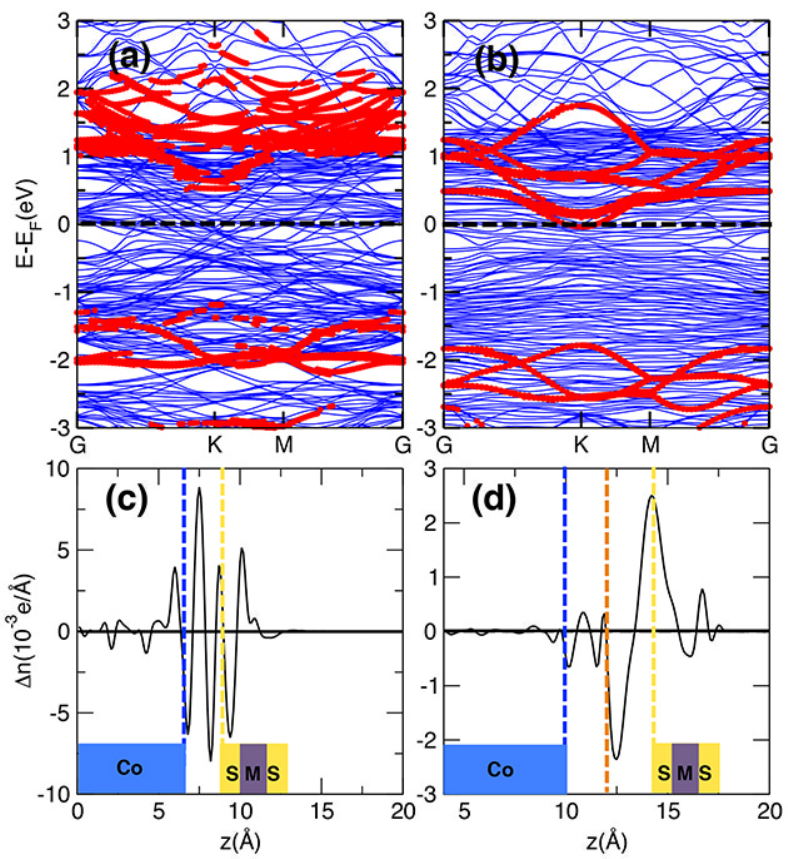}
\caption{(Color online) (a) Band structures of the \SL{}$|$Co(111) and (b) the \SL{}$|$\MBN{}$|$Co(111) interfaces. The red color measures a projection of the wave function on the \MS{} orbitals. The Fermi level is set at zero energy. (c,d) The corresponding electron difference density $\Delta n(z)$. The orange line indicates the position of the \BN{} layer.
}
\label{fig:bnNi}
\end{figure}

The difference between adsorbing \MS{} directly onto a clean Co surface and onto a \MBN{} covered Co surface is observed in the corresponding electronic structures shown in Fig. \ref{fig:bnNi}. Direct adsorption onto Co(111) perturbs the \SL{} bands considerably, due to the interaction at the interface, as shown in Fig. \ref{fig:bnNi}(a). In contrast, adsorbing \MS{} onto a \MBN{}$|$Co(111) substrate hardly perturbs the \SL{} bands at all, as demonstrated by Fig. \ref{fig:bnNi}(b), which is a clear indication that the interaction between \BN{} and \MS{} is weak.

Adsorbing \MS{} directly onto Co leads to the Fermi level being well within the \SL{} band gap, which is not surprising as Co is a high work function metal. In contrast, adsorbing \MS{} onto a \MBN{}$|$Co substrate yields a Fermi level that is pinned at the bottom of the \SL{} conduction band. The electron density differences $\Delta n(z)$ shown in Fig. \ref{fig:bnNi}(c) and \ref{fig:bnNi}(d) substantiate this picture. Figure \ref{fig:bnNi}(c) shows the complicated pattern that is typically associated with a direct \MS{}$|$metal interaction, compare to Fig. \ref{fig:mgbands}(c). $\Delta n(z)$ for \SL{} on a \MBN{}$|$Co shows an accumulation of electrons at the position of \MS{}, and an electron depletion at the position of the \MBN , which reflects an electron transfer from the substrate to \MS{} to equilibrate the Fermi level, compare Figs \ref{fig:bnNi}(d) and \ref{fig:mgbands}(d). 

{\color{red}\it Conclusions.}
We have shown that contacting \SL{} with low work function metals leads to Fermi level pinning at 0.1-0.3 eV below the conduction band edge. This behavior results from the interaction at the interface between the metal and the \MS{}, creating a considerable density of interface states just below the \MS{} conduction band. Inserting a boron nitride (BN) monolayer between the metal and the \MS{} destroys these interface states, and recovers the unperturbed \MS{} band structure. In addition, absorbing \MBN{} on Co(111) or Ni(111) decreases the metal work function by close to 2 eV. We predict that contacting \SL{} with \MBN{}$|$Co or Ni(111) does not give a Schottky barrier. 

{\color{red}\it Acknowledgement.} We thank Menno Bokdam for useful discussions. This work is part of the research program of the Foundation for Fundamental Research on Matter (FOM), which is part of the Netherlands Organisation for Scientific Research (NWO). The use of supercomputer facilities was sponsored by the Physical Sciences Division (EW) of NWO.

%

\end{document}